# A New Line Defect in NdTiO$_3$ Perovskite

Jong Seok Jeong[†,*], Mehmet Topsakal[†], Peng Xu, Bharat Jalan, Renata M. Wentzcovitch, and K. Andre Mkhoyan[*]

[†] These authors contributed equally to this work.

Department of Chemical Engineering and Materials Science, University of Minnesota, Minneapolis, Minnesota 55455, United States

[*] **Corresponding authors:** jsjeong@umn.edu (JSJ); mkhoyan@umn.edu (KAM)

**ABSTRACT:**

Perovskite oxides form an eclectic class of materials owing to their structural flexibility in accommodating cations of different sizes and valences. They host well known point and planar defects, but so far no line defect has been identified other than dislocations. Using analytical scanning transmission electron microscopy (STEM) and *ab initio* calculations we have detected and characterized the atomic and electronic structures of a novel line defect in NdTiO$_3$ perovskite. It appears in STEM images as a perovskite cell rotated by 45 degrees. It consists of self-organized Ti-O vacancy lines replaced by Nd columns surrounding a central Ti-O octahedral chain containing Ti$^{4+}$ ions, as opposed to Ti$^{3+}$ in the host. The distinct Ti valence in this line defect introduces the possibility of engineering exotic conducting properties in a single preferred direction and tailoring novel desirable functionalities in this Mott insulator.

Perovskite oxides form an abundant class of natural materials owing to the plenty of oxygen and the flexibility of the perovskite structure to accommodate cations of different sizes and valences. Remarkable properties have been demonstrated in these materials in bulk, interfaces, or heterostructures, *e.g.*, room-temperature ferroelectricity[1, 2], giant piezoelectricity[3], quantum oscillation[4, 5], two-dimensional superconductivity[6], and pressure-induced spin crossover[7], to mention a few. The chemical diversity, stability under off-stoichiometry, and variety of crystal symmetries in the perovskite structure make it a natural host for a range of defects as well. The possibility of growing these crystals with variable and controlled composition on substrates with desirable interfacial strain allows deliberate incorporation of possibly new defects in these materials[8, 9]. While several unique point and planar defects have been reported in these systems[10-14], no new type of line defect other than dislocations has been observed yet.

The role played by line defects in the basic properties of materials cannot be overstated. Crystalline defects are usually classified in four basic groups, point, line, planar, and bulk defects. The line defect group has only a few members, dominated by three elementary (or topological) defects: edge and screw dislocations and disclinations[15-17]. Despite being conceptually introduced along with dislocations[18], disclinations were experimentally observed only relatively recently[19, 20]. Interestingly, in perovskite with its rich complexity and flexibility of the crystal structure, to the best of our knowledge, no other elementary of even secondary line defects have been reported besides dislocations. Here, we report the discovery of a complex line defect associated with Ti-deficiency in $NdTiO_3$ (NTO) films grown by molecular beam epitaxy (MBE) on a $SrTiO_3$ (STO) substrate. We characterize this defect using a combination of high-resolution analytical scanning transmission electron microscopy (STEM) imaging, atomic-scale spectroscopy, and *ab initio* calculations.

Bulk NTO is a Mott insulator[21] with orthorhombic symmetry (*Pbnm*) and lattice parameters $a_0 = 5.525$ Å, $b_0 = 5.659$ Å, and $c_0 = 7.791$ Å[22]. It can be grown epitaxially as a thin film or as layers in a heterostructure. High-quality stoichiometric NTO films can be grown on STO or $(La_{0.3}Sr_{0.7})(Al_{0.65}Ta_{0.35})O_3$ (LSAT) substrates using hybrid MBE[23]. When analyzing epitaxially grown NTO on STO substrates, it is useful to consider, instead of bulk lattice parameters, tetragonal lattice parameters where $a = b = d_{110} = 3.953$ Å and $c = d_{002} = 3.896$ Å because the tetragonal unit of the NTO lattice is analogous to the STO cubic lattice with $a_0 = 3.905$ Å. Here, we study NTO/STO superlattice structures (consisting of three 6-nm-thick NTO films sandwiched between 3-nm-thick STO films) grown on STO (001) substrates using hybrid MBE. The lattice mismatch results in NTO films with expected biaxial strain: 1.21% compressive in the *a*-axis and 0.23% tensile in the *c*-axis. STEM analysis shows that NTO films tend to grow on STO (001) substrates with their *ac* plane "in-plane" and *b*-direction "out-of-plane".

Two high-angle annular dark-field (HAADF) STEM images of such NTO films are shown in Figs. 1a,b. Viewed along the NTO *c*-axis, the films repeatedly displayed these defects running along the *c*-axis. Missing Ti-O lines are replaced by Nd columns appearing as a perovskite cell rotated by 45 degrees (see Fig. 1b,d,e). When viewed along the NTO *a*-axis, however, such defects were not observed (see Fig. 1a). The preferred direction of the line defect is made clear by the distinct appearance of the NTO crystal structure along the *a*- and *c*-axes (see Figs. 1a,b and *Supporting Information* Fig. S1). Viewed along the *a*-axis, NTO films show a zigzag pattern of Nd columns projected in the *bc* plane, whereas along the *c*-axis NTO shows straight and slightly ellipsoidal Nd atomic columns (Fig. 1c). The presence of Ti-O vacancy lines suggests the formation of this defect could be associated with Ti deficiency in the film, though not intentionally produced here. Alternatively, it may correspond to an equilibrium state achieved under strain and facilitated by the multi-valence nature of Ti, as discussed further below.

The STEM energy dispersive X-ray spectroscopy (EDX) spectrum-image obtained from one such defect shows insignificant Ti signal in the nominal Ti/O column positions, now filled with Nd columns rotated by 45 degrees (Fig. 1e) around the defect core, confirming observations based on the HAADF-STEM imaging. As shown in Fig. 1b, the defects run through the entire thickness of 40-50-nm-thick specimens, whereas it appears that some of them extend only through a part of the specimen (see *Supporting Information* Fig. S2), which are not necessarily an indication of its termination inside the crystal. More discussion about this will be provided later.

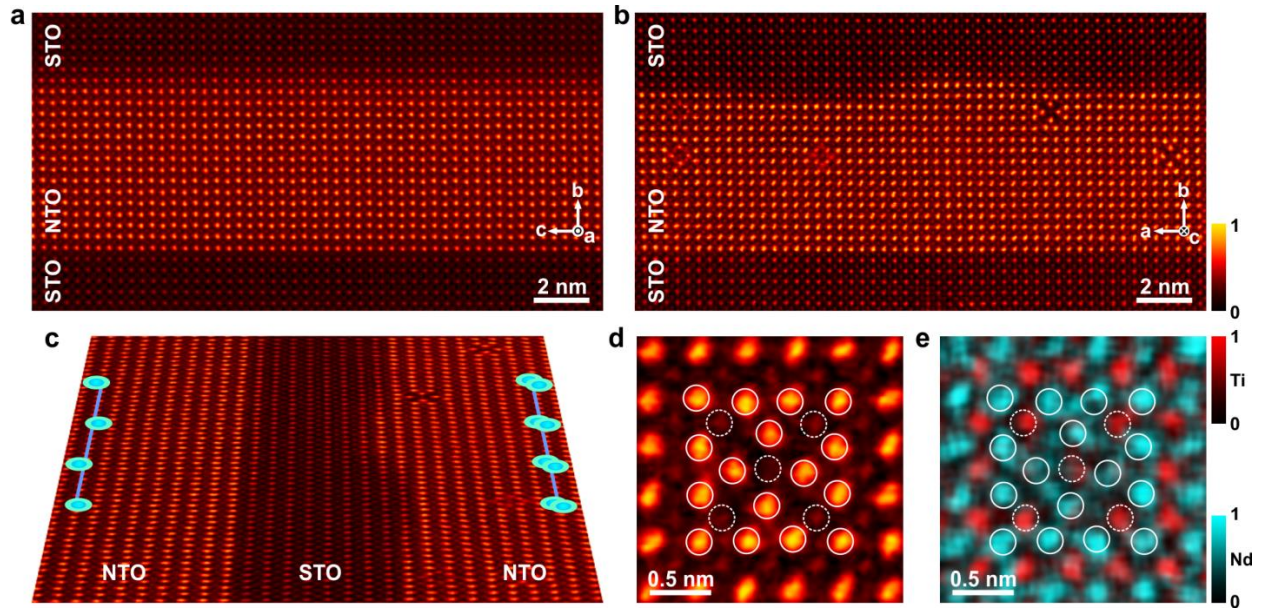

**Figure 1. STEM images of NTO films with and without line defects, grown on an STO (001) substrate.**
**a,** HAADF-STEM image of an NTO film viewed along the *a*-axis. **b,** Another NTO film viewed along the *c*-axis showing the presence of several line defects. **c,** An inclined HAADF image including both films along the *a*- and *c*-axis, is presented to show the arrangement of Nd columns in the NTO films. The left NTO film, shown in **a**, has a zigzag pattern of Nd columns, whereas the right NTO film, shown in **b**, does not. A glimpse of the lattice relaxation of the host NTO around the defects can be seen in this image. A magnified HAADF-STEM image (**d**) is shown along with a STEM-EDX spectrum-image (**e**) of a line defect composed by Nd *L* (cyan) and Ti *L* (red) signals. Positions of Nd and Ti/O columns are indicated with the solid and dotted circles, respectively.

The energetics, stability, and atomic structure of the line defect have been investigated by means of DFT+U calculations (see Methods). Calculations confirmed that a line defect running along the $c$-axis in an NTO crystal is indeed the stable configuration compared to other possible arrangements (Fig. 2). Structural optimizations indicate that in addition to the rotation of Nd columns, there should be a half-unit-cell displacement of the entire Ti-O-Nd core along the $c$-axis (see *Supporting Information* Movie S1). Such shift is undetectable in the HAADF-STEM images recorded along the $c$-axis but lowers the total energy by ~8.5 eV/layer compared to the unshifted defect (configuration LD' in Fig. 2d). The stability of the line defect structure (configuration LD in Fig. 2c) was also tested on the rotation of the central Ti-O octahedral chain (see *Supporting Information* Movie S2), indicating that it is primarily the Nd columns that rotate around the central octahedral chain after the defect is fully relaxed. Compared to isolated and randomly arranged Ti-O vacancy pairs (configuration VRand in Fig. 2a), a configuration that could pre-exist their organization, the line defect is favored by approximately ~5 eV/layer with 4 Ti-O pairs each. It should be noted that, based on several inspected cases, the energy spread among various random configurations of interacting Ti-O vacancy pairs is within 0.5 eV/layer. Vacancy concentration in these calculations corresponds to $NdTi_{1-x}O_{3-x}$ with $x = 0.111$. The system with the same concentration of Ti-O vacancies but far apart to be considered isolated has even higher energy (see *Supporting Information* Fig. S3). The line defect is also energetically favorable compared to a defect in which the central octahedral chain plus surrounding Nd columns shift by a half unit cell along the $c$-axis but no rotation of Nd columns is allowed (configuration VOrd' in Fig. 2d). For completeness of analysis, calculations were performed with both experimental lattice parameters ($a = 3.90(1)$ Å, $b = 3.933(1)$ Å, and $c = 3.91(1)$ Å) obtained by a combination of high-resolution STEM imaging and X-ray diffraction using the STO lattice as a reference, and unstrained *ab initio* lattice parameters at zero pressure. It should be noted that the measurements indicate biaxial compression in the *ab* plane and tension in the *c*-direction in the NTO film (see Methods). The phenomenological outcome of both calculations is the same, as shown in Figs. 2d,e. The stability of the defect without strain suggests that, similar to dislocations, once created this defect will remain stable even after strain is removed.

Calculations were also performed for a line defect running along the *a*-axis and results tell the same story (see Figs. 2d,e): the line defect is energetically more favorable than other configurations. The defects running along the *a*-axis have similar energy and should be as common as those running along the *c*-axis in unstrained NTO. However, when biaxial strain is present as in our film, the line defects running along the *c*-axis are approximately ~0.5 eV/layer more stable than those running along the *a*-axis. Recall that along the *a*-axis the NTO film experiences considerable compressive strain (see Methods). The line defect with missing neighboring Ti-O columns is slightly "loose", which helps alleviating substrate imposed stresses. Careful analysis of the HAADF-STEM images and the DFT+U structure show that relaxation of the host NTO structure around a line defect indeed occurs (see *Supporting Information* Fig. S4), further supporting the structural identification of these defects. The detailed atomic structure of the defect and its simulated HAADF-STEM image are shown in *Supporting Information* Figs. S5 and S6, respectively.

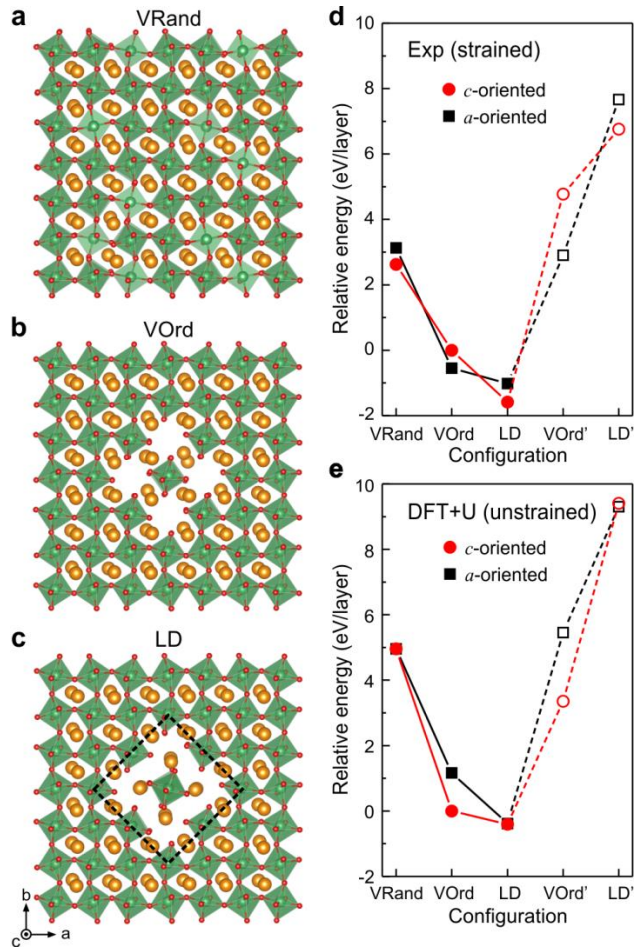

**Figure 2. (DFT+U)-calculated structures and relative energy comparison of possible Ti/O-deficient defects in NTO. a,** Atomic structure of NTO with a random distribution of Ti-O vacancies (VRand). **b,** Possible structure of NTO with ordered Ti-O vacancies (VOrd), in which they surround one Ti-O-Nd unit. **c,** Calculated structure of the line defect (LD) including a half-unit-cell shift along the *c*-axis. Yellow spheres represent Nd atoms, and green octahedra are Ti atoms coordinated by six O atoms (red spheres). **d-e,** Relative energies of structures presented in **a-c**, calculated using the experimentally determined, i.e., strained (**d**) and (DFT+U) predicted (unstrained) (**e**) lattice parameters. Structure VOrd is used as a reference state. Two additional structures (not shown) were also calculated and compared: a structure in which the center Ti-O-Nd unit in VOrd shifts by a half unit cell along the central axis but is not allowed to rotate (VOrd') and similarly a structure in which the center Ti-O-Nd unit in LD shifts by a half unit cell along the central axis but is not allowed to rotate (LD'). In these calculations, defects running along both directions, the *c*-axis (*c*-oriented) and *a*-axis (*a*-oriented), are considered. The energies of the VOrd structures running along the *c*-axis are chosen as a reference. Because both VOrd' and LD' structures relax into LD configuration, their atomic positions are fixed in order to calculate their total energies (presented by the open symbols and dotted lines).

At this point, it is not clear whether the formation of Ti-O vacancies is the result of partial strain release in the film facilitated by the multivalence of Ti or due to small and local compositional variation during film growth. Vacancies are always expected to exist in any material at finite temperatures. Transition metal oxides are notorious for having metal or oxygen vacancies, or both simultaneously, or for forming several compounds with various metal to oxygen ratios. This is particularly true for the Ti-O system that exists as $TiO$, $Ti_2O_3$, and $TiO_2$. Besides, TiO exists in a rocksalt-related structure consisting of ordered Ti-O vacancies (as in $Ti_5O_5$)[24]. This also occurs with other transition metal oxides, e.g., rocksalt-type NbO (as in $Nb_3O_3$) where 25% NbO vacancies align along the cube diagonal[25]. Therefore, the organization of TiO vacancies along a line in $NdTiO_3$ films does not seem too surprising. Besides the Nd-Ti-O system forms more than one stoichiometric compound. Particularly relevant is the $Nd_2TiO_5$ system (with $Ti^{4+}$), which can be viewed as a 50-50% intermediate compound between $Nd_2O_3$ and $TiO_2$[26]. Conversely, $Nd_2TiO_5$ and TiO might be viewed as dissociation products of $NdTiO_3$. Though phase relations among $NdTiO_3$, $Nd_2TiO_5$, $TiO_2$, and TiO are not known, the existence of a stable compound with stoichiometry $Nd_2TiO_5$, though with different structure, is very relevant because the line defect core, i.e., the region surrounded by the fine-dashed black line in Fig. 2c has precisely this composition. Therefore, this new line defect can be a multi-faceted phenomenon. The basic requirement is the presence of (i) a multivalent cation; (ii) transition metal off-stoichiometry during growth may facilitate its formation, but might not be required if the defect corresponds to a thermodynamic equilibrium state forced by (iii) strain or lattice mismatch. The presence of this defect in unstrained bulk is not ruled out, as indicated in Fig. 2e, but strain stabilizes it in one preferred direction, as seen in Fig. 2d.

It should be noted that HAADF-STEM images and EDX maps are not sensitive to the exact O content in the structure and the defect could be *n*-doped with further O vacancies at or near the line defect. To evaluate the stability of the defect with respect to O content, additional *ab initio* calculations including one extra O vacancy in the core of the defect were performed. The structural optimization produced an almost identical line defect (see *Supporting Information* Fig. S7). However, as discussed below, this ambiguity is addressed to a certain extent using EELS measurements to test the *ab initio* prediction that Ti ions in the line defect core are $Ti^{4+}$.

DFT+U results confirming and clarifying the atomic structure of the line defect were further analyzed to understand its electronic structure. First, the defect structure indicates that Ti ions in the defect core should be $Ti^{4+}$, in contrast to $Ti^{3+}$ in bulk NTO (Figs. 3a,b). The charge density ($\rho\uparrow + \rho\downarrow$) and charge polarization density ($\rho\uparrow - \rho\downarrow$) shown in Fig. 3b are consistent with this interpretation. To test this prediction, site-specific electron energy-loss spectroscopy (EELS) measurements were performed using STEM (see *Supporting Information* Fig. S8). Element-specific core-level EELS measurements provide spectra with a detailed fine structure that often can be directly compared with the corresponding partial density of states (p-DOS) from that element[27-29]. O *K*–edge and Ti $L_{2,3}$–edge core-level EELS spectra from on and off the defect sites were recorded (Figs. 3c,d). The fine structure of the O *K*–edge is directly compared with the calculated O 2*p* p-DOS in the conduction band. As shown in Fig. 3c, the agreement is quite good. Similar analysis of the Ti core-level EELS $L_{2,3}$–edge is not straightforward, as it includes both the 3*s* and 3*d* p-DOS from Ti and the initial 2*p* core level in split in two levels, $2p^{1/2}$ and $2p^{3/2}$, produced by spin-orbit interaction. More sophisticated methods going beyond the independent particle approximation are required for these calculations[30]. However, Ti $L_{2,3}$–edge spectra are known to be quite distinct for $Ti^{4+}$ or $Ti^{3+}$ in perovskite oxides[31-34]. The Ti $L_{2,3}$–edge EELS spectrum recorded from a defect site shows $Ti^{4+}$ features with strong peaks at 459.0, 461.1, 464.7, and 466.3 eV, whereas a spectrum recorded off a defect site, as expected, shows $Ti^{3+}$ character (Fig. 3d).

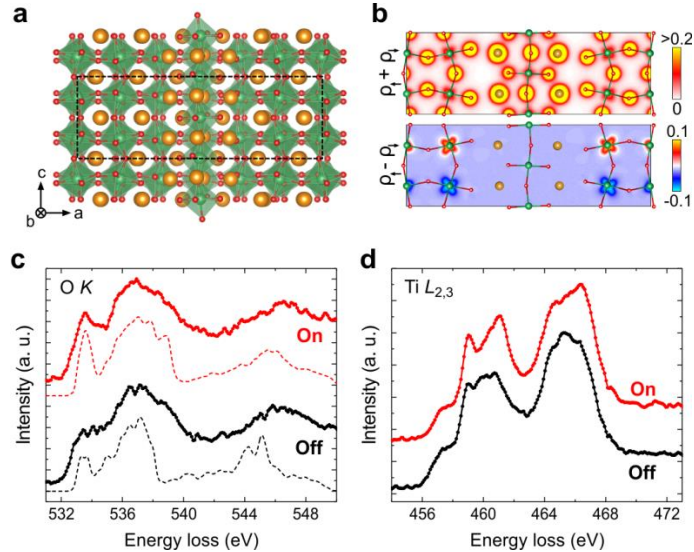

**Figure 3. Electronic structure of a line defect in NTO. a,** (DFT+U)-calculated atomic structure of a line defect that runs along the *c*-axis and is viewed along the *b*-axis. Yellow spheres represent Nd atoms, and green octahedra correspond to Ti atoms surrounded by six O atoms (red spheres). **b,** Charge densities from the plane passing through the defect core indicated in **a**. The tetravalent nature of Ti atoms in the defect is clearly visible when a $\rho_\uparrow - \rho_\downarrow$ charge density map is compared with a $\rho_\uparrow + \rho_\downarrow$ map (see also *Supporting Information* Fig. S9). Here, an atomic model of the structure is overlaid onto the charge density maps for clarity. **c,** Measured, on and off defect (see *Supporting Information* Fig. S8), O *K*–edge EELS data with its fine structures (bold lines) directly compared to calculated O 2*p*-DOS (dotted lines). **d,** Measured, on and off defect, Ti $L_{2,3}$–edge EELS fine structures showing the $Ti^{4+}$ feature when it is measured at the defect site and the $Ti^{3+}$ character from the NTO host.

The result of detailed EELS analysis of the defect core is presented in Fig. 4. EELS data recorded across the core of the defect show distinct changes in its fine structure, Fig. 4b. Ti $L_{2,3}$ spectra show that $Ti^{3+}$ character changes to predominantly $Ti^{4+}$ (with distinct crystal field and intensity ratio between $e_g$ and $t_{2g}$ peaks) when the probe moves from the bulk to the core of the defect, even though the transition is not abrupt and occurs over a distance of 1 nm from the defect center (see also *Supporting Information* Fig. S9). O *K* spectra also show changes in the fine structure across the defect. The changes are in quantitative agreement with the O 2$p$ p-DOS obtained from DFT+U calculations in terms of the edge onset, onset peak height, and width of the second peak at 535-540 eV (Figs. 4d,e). It appears that the main differences between electronic states of Ti and O atoms occur at the boundary between defect and the bulk, and not at its core. These results indicate that our *ab initio* structural determination is quite reliable and the electronic structure of the line defect is indeed considerably different from that of bulk NdTiO$_3$.

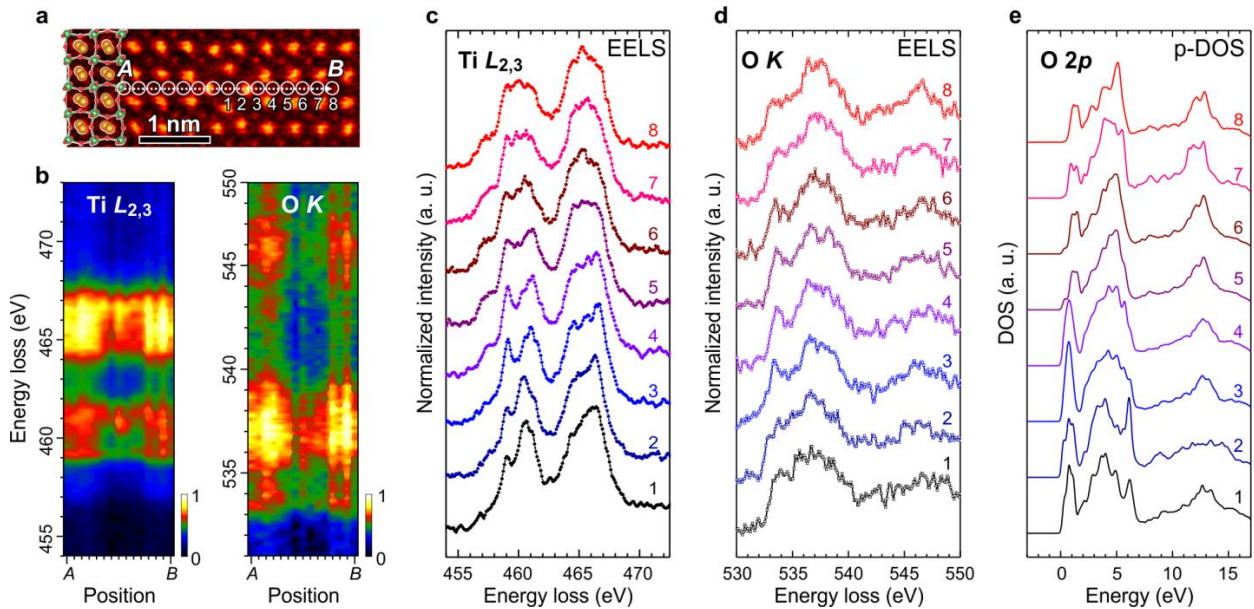

**Figure 4. Detailed EELS and DFT analysis of the defect core. a,** ADF-STEM image of a line defect showing atomic structure and the positions of EELS line profile (from *A* to *B*). Superimposed circles represent probe positions in the EELS line scan. The image was low-pass-filtered to 1.67 Å$^{-1}$. **b,** EELS Ti $L_{2,3}$ (left panel) and O $K$ (right panel) spectra vs. position. The EELS data images are separately normalized. **c-d,** The fine structure of the Ti $L_{2,3}$ and O $K$ measured across the defect. **e,** (DFT+U)-calculated O 2$p$ p-DOS at each position. Here Gaussian broadening with full width at half maximum of 0.3 eV was applied to calculated p-DOS. Probe positions (1-8) correspond to those labeled in **a**.

The electronic structure of NTO with a line defect is also strikingly different from that of bulk NTO (see Fig. 5) and in the presence of a line defect, the gap narrows by approximately ~0.4 eV. The defect band structure shown in Fig. 5 clearly indicates that only along the defect line, Γ-Z line in the Brillouin Zone, excited carriers have finite mobility. Along other directions, states within ~0.5 eV from the conduction band bottom are all quite flat, pointing to the resilient one-dimensional (1D) nature of the emergent conductivity in the *n*-doped system, even at very high temperatures. Therefore, if n-doped, the new line defect containing $Ti^{4+}$ might lead to the emergence of exotic 1D conductance in this Mott insulator. The nature of the conductivity should be very sensitive to dopant concentration, varying from possibly polaronic conductivity at small dopant concentrations, to possibly metallic conductivity at intermediate dopant concentrations, back to Mott insulating behavior at one $e^-$ per $Ti^{4+}$ doping level. For emergence of 1D conductivity, single domains with all line defects oriented in parallel should be engineered. As shown here, this might be accomplished using substrate induced strains.

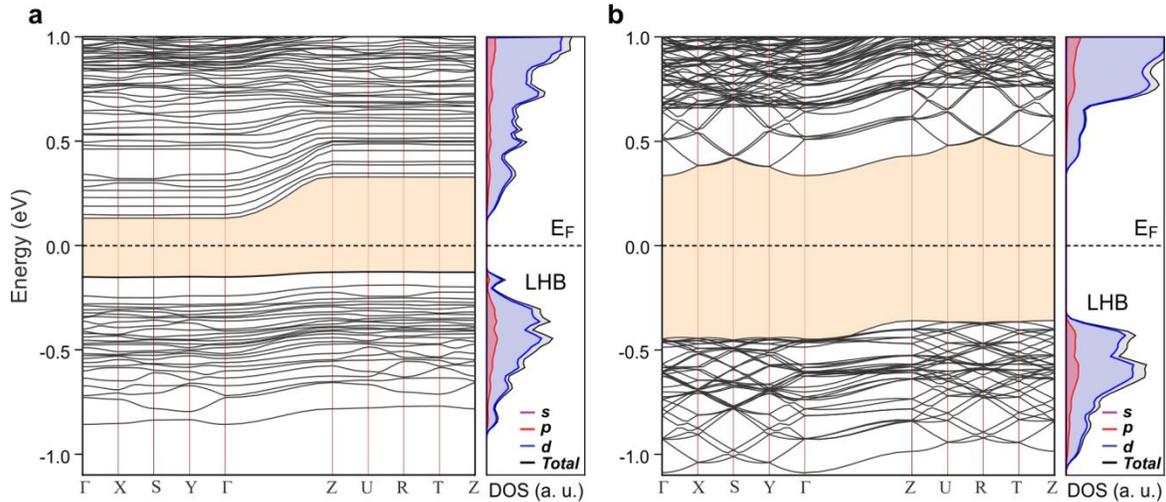

**Figure 5. Band structures of NTO.** Band structures with DOS of the NTO crystal **a**, with and **b**, without a line defect. The Fermi level ($E_F$) is set in the middle of the band gap (shaded). Valence bands within 1 eV from $E_F$ constitute the filled Ti-3*d* lower Hubbard bands (LHB).

Analysis of the HAADF-STEM images, based on the orientation of neighboring Nd atomic dumbbells, indicates that there are two different types of line defects: right-handed and left-handed (*Supporting Information* Fig. S10). The right- or left-handedness of the line defect is essentially produced by a negative or positive rotation of Nd columns in the NTO lattice. Alternatively, they can be viewed as the same defect but shifted by half of a unit cell in the plane, perpendicular to a defect line, i.e., along the *a*- or *b*-axes (Fig. S10b,d). These two cases are crystallographically distinct, as there is no combination of symmetry operations that will transform one into the other. DFT+U results show no significant differences between the electronic structures of right- and left-handed defects. It should be noted that such differentiation between the right- and left-handed line defects is only valid under biaxial strain in the (010) *ac*-plane, as is the case here. In the absence of strain, right- and left-handed defects are crystallographically equivalent, and one can be transformed into the other by combining mirror operations and a 90-degree rotation about its axis, the *c*-axis in this case.

The three factors seemingly responsible for this defect formation, i.e., presence of a multivalent ion, slightly off stoichiometry (specifically deficiency of a multivalent ion) and substrate induced strain, co-exist commonly in perovskite heterostructures or epitaxially grown films of perovskites (e.g., titanates, vanadates, niobates, manganites, ferrites, etc). We expect, therefore, that such line defects might be able to form in several of these perovskites, though the detailed nature of the substrate induced strain, film thickness, etc. might be relevant for its stabilization. Indeed, a defect with similar appearance in ADF-STEM images has been observed in $Bi(Mn,Fe)O_3$ films, consistent with our expectation[35].

Finally, the question here is whether this line defect is an elementary defect, like dislocations and disclinations, or a secondary defect, which can be reduced to a combination of existing elementary defects. While at this point we cannot conclusively classify to which category this defect belongs to, we would like to provide initial discussion about this defect for future detailed study. The primary factors that make a line defect elementary are: (i) it should be induced by macroscopic deformation of a crystal with a characteristic vector or an angle defined by the lattice parameters (Burgers vector for dislocations or deficit angle for disclinations) and (ii) it should not end inside the crystal[36, 37]. One-dimensional stresses in the crystal produce dislocations (edge and screw), whereas two-dimensional (2D) stresses influence the formation of this new line defect. The macroscopic 2D strain in the crystal can be accommodated by simply shifting the central unit or the "core" line by a half unit cell along the core line direction. Atomic rearrangement of the core minimizes strain and the overall energy. Additional calculations indicate that 2D compression, applied to the plane perpendicular to the line defect, energetically favors the formation of this line defect (*Supporting Information* Fig. S11). As with dislocations and disclinations, if the line defect is elementary, it is expected to see climbs and bends as they cannot end inside the crystal. We often observed pairs of *partial* defects that could be climbs visualized along the defect line (*Supporting Information* Fig. S2). Bends, on the other hand, are not easy to detect in atomic-resolution HAADF-STEM images, as a portion of the defect will run non-parallel to the incident electron beam. They will appear simply as individual partial defects. *Supporting Information* Fig. S12 provides a visual summary and comparison with other elementary defects. However, we cannot rule out a possibility of it being a secondary defect. Classification of this line defect is beyond the scope of this work.

In conclusion, we report the discovery of a new line defect in strained NTO perovskite in NTO/STO heterostructures. The defect was detected using aberration-corrected STEM and characterized with assistance of *ab initio* calculations. EELS measurements and DFT+U results indicate that the oxidation state of Ti in the defect and its associated electronic properties differ from those of the host crystal. In addition to contributing a new member to the limited group of line defects, these findings introduce new opportunities in the field of perovskite oxides. This new line defect could also be a building block for grain boundaries in perovskites, similar to those in metals represented by an array of dislocations. We expect that by controlling strain in the crystal, the number of line defects and the electronic properties of the crystal could potentially be manipulated. At this stage it appears that conductivity along a strongly preferred direction could emerge upon *n*-doping, e.g., by manipulating oxygen fugacity during growth, or at high temperatures. Considering the wide variety of multi-valent perovskite oxides that can be produced using existing growth technologies, this new line defect might be created, with the proper control of growth parameters and compositions, in others as well.

**MATERIALS AND METHODS:**

**MBE growth of NTO films.** NTO (6 nm)/STO (3 nm) heterostructures were grown on STO (001) substrates (Crystech GmbH, Germany) using a hybrid MBE approach[23], which employs titanium tetra-isopropoxide (TTIP) (99.999% from Sigma-Aldrich, USA) as the metal-organic precursor for Ti and solid elemental sources for Nd (99.99% from Ames Lab, USA) and Sr (99.99% from Sigma-Aldrich, USA). No additional oxygen was used because TTIP also supplies oxygen. TTIP precursor was introduced via a gas injector (E-Science Inc., USA) by thermal evaporation in a bubbler, which was connected to the gas injector through a gas inlet system equipped with a linear leak valve and a baratron. TTIP flux was controlled precisely by regulating the linear leak valve using a feedback contol. Given the high volatility of the TTIP precursor at relatively low bubbler temperature, no carrier gas was used. Phase pure, epitaxial NTO films were grown at a substrate temperature of 900 °C by controlling the Nd/Ti beam equivalent pressure. Likewise, the cation stoichiometry of STO layers was optimized by controlling the Sr/Ti beam equivalent pressure ratio during MBE growth. The films investigated in this study were grown in nominally close to optimized growth conditions for cation stoichiometry.

**Lattice parameter measurement.** Experimental lattice parameters were estimated to be $a = 3.90(1)$ Å, $b = 3.933(1)$ Å, and $c = 3.91(1)$ Å by a combination of high-resolution STEM imaging and X-ray diffraction using STO lattice as a reference. They indicate the presence of biaxial compression in the $ab$ plane ($-1.34 \pm 0.25\%$ and $-0.51 \pm 0.03\%$ in the $a$- and $b$-directions, respectively) and tension in the $c$-direction ($0.36 \pm 0.26\%$) in the NTO film, indicating strain compared to bulk NTO.

**STEM specimen preparation.** Cross-sectional STEM specimens were prepared by a focused ion beam (FIB, FEI Quanta 200 3D) using 30 kV Ga ions, followed by 5 kV ion milling to remove damaged layers on the specimen surfaces. Then, electron-transparent and relatively damage-free specimens were prepared by additional Ar-ion milling (Fischione ion miller, Model 1010) for less than 5 min with a 1.5–2.5 kV beam voltage and a 6° incident beam angle.

**STEM operation and HAADF-STEM imaging.** STEM data were obtained using an aberration-corrected and monochromated FEI Titan G2 60-300 STEM. EELS spectra were recorded using a Gatan Enfinium ER spectrometer attached to the STEM with an energy resolution of 0.14 eV that was achieved with a 0.01 eV/channel energy dispersion. EDX spectrum-imaging was performed using a Super-X system also attached to the STEM. Two different experimental conditions were used for the HAADF-STEM imaging and spectroscopy: 300 keV with the monochromator inactivated for the HAADF imaging and EDX spectroscopy, and 200 keV with the monochromator activated for the EELS spectroscopy. The convergence semi-angles of the STEM incident beam were 24.5 mrad and 17.3 mrad during the 300 keV and 200 keV–mono operation, respectively, with corresponding HAADF detector inner angles of 49.5 mrad and 53.2 mrad. The electron probe was corrected using a CEOS-DCOR probe corrector. The spatial resolution was measured to be (i) ~0.8 Å for the HAADF-STEM imaging at 300 keV with a probe current of ~55 pA, and (ii) ~1.1 Å during EELS data collection at 200 keV with a probe current of ~58 pA. The spatial resolution of the STEM was determined by analysis of the fast-Fourier-transform (FFT) of high-resolution HAADF-STEM images of a standard test sample, which was a carbon replica with Au waffle-pattern gratings. For the HAADF-STEM image acquisition, 512×512 pixel$^2$ and 2048×2048 pixel$^2$ scanning window sizes were used for the search and acquisition modes, respectively. Dwell times of 2–6 µs/pixel were used for both scans.

**HAADF-STEM image processing.** The HAADF-STEM images shown in the main text are low-pass filtered, limiting the information below 0.65 Å. The HAADF-STEM images in Fig. 5 were further processed by averaging several images using a cross-correlation algorithm. The image in Fig. 5a is the cross-correlated average of nine defect images, whereas the image in Fig. 5b is the average of only three, which explains the differences in the signal-to-noise ratios seen in these images. The images in *Supporting Information* Figs. S1c,d are the cross-correlated averages of fifteen and fourteen individual images, respectively.

**EDX spectrum-imaging.** Regions with a defect were selected using a HAADF-STEM image, and EDX spectrum-imaging was performed with frame-by-frame spatial drift correction enabled using Bruker Esprit 1.9 software. A dwell time of 1 µs with a 100×100 pixel$^2$ scan grid was used with a total acquisition time of 58 s.

**EELS measurements.** STEM-EELS measurements were conducted in "spot" mode with a spectrometer collection angle of 25.8 mrad. The proper probe currents and dwell times were selected to minimize the electron-beam-induced damage and also to minimize the effects of specimen drift. Dual-mode EELS was used to simultaneously acquire both the zero-loss and core-loss EELS spectra from each spot to ensure that the energy drift during spectrum acquisition can be compensated.

**Specimen-thickness measurement.** Measured low-loss EELS spectra were used to estimate specimen thickness ($t$). The mean free paths ($\lambda$) for bulk plasmon generation in NTO under our experimental conditions were determined by comparing EELS data from adjacent STO layers ($\lambda_{STO}$ = 123 nm) at the NTO/STO interfaces and were approximately 110 for 300 and 106 nm 200 keV. The typical thickness of NTO specimens used in this study is 45 ± 6 nm.

*Ab initio* **calculations.** *Ab initio* calculations were performed in the framework of density functional theory with Hubbard U corrections (DFT+U) as implemented in the Vienna *ab initio* Simulation Package (VASP) code[38]. The spin-polarized generalized gradient approximation (GGA) in the Perdew-Burke-Ernzenhof (PBE) parametrization[39] was adopted as the electronic exchange and correlation functional and a Hubbard U = 2.07 eV was added to $Ti^{+3}$ ions. The interaction between ions and valence electrons is described by the projected augmented wave (PAW) method[40] with a plane wave cutoff of 350 eV. Ions in VRand, VOrd, and LD were relaxed until the forces were smaller than 0.005 eV/Å, in which case the total energy was simultaneously minimized. Configurations VOrd' and LD' are not relaxed because they are unstable configurations. To reduce effects of image interactions that arise due to the use of periodic boundary conditions, a large supercell containing 344 atoms ($Nd_{72}Ti_{64}O_{208}$) was selected for these calculations. We removed eight Ti and O atoms to mimic the Ti-O deficient structure. For all structures presented in this study, the Brillouin zone was sampled at a 2×2×4 shifted *k*-point grid using the tetrahedron method[41]. The atomic structures presented in Figs. 2a-c, 3a,b, and 4a are drawn using the VESTA software[42].

**ASSOCIATED CONTENT:**

**Supporting Information.** Additional information about materials and methods, structure, STEM images, DFT+U calculations, and movies.

**AUTHOR INFORMATION:**

**Corresponding authors.** jsjeong@umn.edu (JSJ); mkhoyan@umn.edu (KAM).

**Author Contributions.** J.S.J and M.T. contributed equally to this work. J.S.J and K.A.M. conceived the experiments. J.S.J. performed the experiments and characterization of the materials. M.T. conducted the DFT+U calculations and M.T. and R.M.W. analyzed the results. P.X. and B.J. grew materials. K.A.M. supervised the project. All authors discussed the results and participated in writing the paper.

**Notes.** The authors declare no competing financial interest.

**ACKNOWLEDGMENTS:**

This work was supported in part by NSF MRSEC under awards DMR-0819885 and DMR-1420013, also in part by NSF DMR-1410888, NSF EAR-134866 and EAR-1319361, and by the Defense Threat Reduction Agency, Basic Research Award #HDTRA1-14-1-0042, to the University of Minnesota. Computational resources were partly provided by Blue Waters sustained-petascale computing project, which is supported by the NSF under awards OCI-0725070 and ACI-1238993 and the state of Illinois. Blue Waters is a joint effort of the University of Illinois at Urbana-Champaign and its National Center for Supercomputing Applications. Structural characterization including STEM analysis was performed in the Characterization Facility of the University of Minnesota, which receives partial support from the NSF through the MRSEC program. Multislice computer simulations were performed using resources provided by the Minnesota Supercomputing Institute. The authors acknowledge use of facilities in the Minnesota Nano Center. The authors also thank Drs. Frank Bates and Chris Leighton for helpful discussions, and Drs. Danielle Reifsnyder Hickey and Michael Odlyzko for critically reading the manuscript.